\documentclass[aip,twocolumn,superscriptaddress,
preprintnumbers,
amsmath,amssymb,showkeys,floatfix,nofootinbib,reprint]{revtex4-1}

\usepackage{amsmath}
\usepackage{amsfonts}
\usepackage{amssymb}
\usepackage{graphicx}
\usepackage{color}
\usepackage{bm}
\usepackage{hyperref}

\begin{document}

%
%
% -------------------------------- our notations --------------------------- %
%

\def\ada#1{\textcolor{blue}{#1}}
\def\jonas#1{\textcolor{red}{#1}}

\def\ket#1{ $ \left\vert  #1   \right\rangle $}
\def\ketm#1{  \left\vert  #1   \right\rangle   }
\def\bra#1{ $ \left\langle  #1   \right\vert $ }
\def\bram#1{  \left\langle  #1   \right\vert   }
\def\spr#1#2{ $ \left\langle #1 \left\vert \right. #2 \right\rangle $ }
\def\sprm#1#2{  \left\langle #1 \left\vert \right. #2 \right\rangle   }
\def\me#1#2#3{ $ \left\langle #1 \left\vert  #2 \right\vert #3 \right\rangle $}
\def\mem#1#2#3{  \left\langle #1 \left\vert  #2 \right\vert #3 \right\rangle   }
\def\redme#1#2#3{ $ \left\langle #1 \left\Vert
                  #2 \right\Vert #3 \right\rangle $ }
\def\redmem#1#2#3{  \left\langle #1 \left\Vert
                  #2 \right\Vert #3 \right\rangle   }
\def\threej#1#2#3#4#5#6{ $ \left( \matrix{ #1 & #2 & #3  \cr
                                           #4 & #5 & #6  } \right) $ }
\def\threejm#1#2#3#4#5#6{  \left( \matrix{ #1 & #2 & #3  \cr
                                           #4 & #5 & #6  } \right)   }
\def\sixj#1#2#3#4#5#6{ $ \left\{ \matrix{ #1 & #2 & #3  \cr
                                          #4 & #5 & #6  } \right\} $ }
\def\sixjm#1#2#3#4#5#6{  \left\{ \matrix{ #1 & #2 & #3  \cr
                                          #4 & #5 & #6  } \right\} }

\def\ninejm#1#2#3#4#5#6#7#8#9{  \left\{ \matrix{ #1 & #2 & #3  \cr
                                                 #4 & #5 & #6  \cr
                         #7 & #8 & #9  } \right\}   }
%
%
% ---------------------------- end of our notations --------------------------- %
%
%

%
% -----------------------------------------    Title of paper ---------------------------------------
%

\title{Neutron production from thermonuclear reactions in laser-generated plasmas}

%
% ------------------------------------------   List of authors --------------------------------------
%

\author{Yuanbin \surname{Wu}}
\email{yuanbin.wu@mpi-hd.mpg.de}
\affiliation{Max-Planck-Institut f\"ur Kernphysik, Saupfercheckweg 1, D-69117 Heidelberg, Germany}

\date{\today}

%
%
%
% ---------------------------------------------------- Abstract ---------------------------------------------
%
%
%
%
\begin{abstract}

The production of intense neutron beams via thermonuclear reactions in  laser-generated plasmas is investigated theoretically. So far, state-of-the-art neutron beams are produced via laser-induced particle acceleration leading to high-energy particle beams that subsequently interact with a secondary target. Here we show that neutron beams of two orders of magnitude narrower bandwidth can be obtained from thermonuclear reactions in plasmas generated by Petawatt-class lasers. The intensity of such neutron beams is about one or two orders of magnitude lower than the one of the state-of-the-art laser-driven neutron beams. We study to this end the reaction $^2$H($d$, $n$)$^3$He in plasmas generated by Petawatt-class lasers interacting with D$_2$ gas jet targets and CD$_2$ solid-state targets. The results also shows the possibility of direct measurements of reaction rates at low temperatures of astrophysical interests. In addition, the use of CD$_2$ solid-state targets can also lead to great enhancements on the plasma screening compared to the case of D$_2$ gas jet targets, opening new possibilities to study this so far unsolved issue in the field of astrophysics. 

\end{abstract}
%

%\pacs{
%52.50.Jm, %	Plasma production and heating by laser beams (laser-foil, laser-cluster, etc.)
%25.45.-z, % 2H-induced reactions
%28.20.-v, % Neutron physics (see also 25.40.-h Nucleon-induced reactions and 25.85.Ec Neutron-induced fission)
%29.25.Dz, % Neutron sources
%26.20.-f, % Hydrostatic stellar nucleosynthesis (see also 97.10.Cv Stellar structure, interiors, evolution, nucleosynthesis, ages in astronomy)
%52.59.-f, % Intense particle beams and radiation sources (see also 29.25.-t Particle sources and targets, and 29.27.-a Beams in particle accelerators, in instrumentation for elementary-particle and nuclear physics)
%52.65.Rr, %	Particle-in-cell method
%52.72.+v, % Laboratory studies of space- and astrophysical-plasma processes (see also 94.05.Rx in space plasma physics)
%97.10.Cv, % Stellar structure, interiors, evolution, nucleosynthesis, ages
%}

%\keywords{}

\maketitle

%%%--------------------------------------introduction-------------------------------------------------------
\section{Introduction}

Thermonuclear reactions occur in plasma environments in which the thermal energy of the ions can overcome the electrostatic repulsion in a collision between nuclei, leading to nuclear reactions \cite{AtzeniBook2004}. The development of laser technology in the past decades provides a powerful tool for the study of nuclear reactions in laser-generated plasmas. Lasers provide the opportunity to access plasma parameter regimes which cannot be accessed in accelerator-based experiments, such as direct measurements of nuclear reactions in nucleosynthesis-relevant energies and plasma effects on nuclear reactions \cite{NegoitaRRP2016, CaseyPRL2012, ZylstraPRL2016, CerjanJPG2018}, and might thus significantly advance our understanding in astrophysics. On the other hand, industrial applications of such studies, such as laser-induced ignition which may provide a future source of alternative energy \cite{CerjanJPG2018, HurricaneNature2014, OlsonPRL2016}, have also attracted a great deal of attention.

Neutron production is one of the key areas of the field of nuclear reactions in laser-generated plasmas \cite{MaPRL2014, DoppnerPRL2015, DitmireNature1999, HigginsonPRL2015, Roth2013PRL, Pomerantz2014PRL}. Normally the experimental access to high neutron flux is mainly at large-scale reactor and accelerator-based facilities. However, the development of Petawatt-class lasers provides the opportunity of having intense neutron beams generated by comparatively smaller-scale laser facilities \cite{HigginsonPRL2015, Roth2013PRL, Pomerantz2014PRL}. The common solution for neutron production in Petawatt-class laser facilities is via high-energy particle beams interacting with a target (beam-target interaction), in which lasers are used for the acceleration of particles. In this manner, intense neutron beams with the order of $10^9$-$10^{10}$ per pulse can be obtained \cite{HigginsonPRL2015, Roth2013PRL, Pomerantz2014PRL}.

In this article, we study intense neutron beams produced from thermonuclear reactions in laser-generated plasmas. We analyze the reaction $^2$H($d$, $n$)$^3$He in plasmas generated by Petawatt-class lasers interacting with D$_2$ gas jet targets and CD$_2$ solid-state targets. Intense neutron beams with narrow bandwidth can be obtained from thermonuclear reactions in plasmas generated by Petawatt-class lasers, which are about two orders of magnitude narrower in the neutron-energy bandwidth compared to today's state-of-the-art laser-driven neutron beams \cite{Roth2013PRL}. The intensity of such neutron beams is about one or two orders of magnitude lower than the one of the state-of-the-art laser-driven neutron beams \cite{Roth2013PRL}. Such intense neutron beams with narrow bandwidth have numerous applications in both industry and fundamental research, such as the interrogation of material and life science \cite{ZaccaiScience2000, MaNatNano2013}, nuclear fission and fusion research \cite{PerkinsNF2000}, and neutron capture experiments for fundamental nuclear physics and nuclear astrophysics \cite{BleuelPFR2016}. On the other hand for the low temperature regime, the results show the possibility of direct measurements of reaction rates at low temperatures of astrophysical interests \cite{Adelberger2011RMP, NegoitaRRP2016}. The use of CD$_2$ solid-state targets can also lead to great enhancements on the plasma screening compared to the case of D$_2$ gas jet targets, offering the possibility to access to this so far unsolved issue in astrophysics \cite{Adelberger2011RMP, NegoitaRRP2016, CerjanJPG2018}. 

We note that in order to obtain nuclear reaction rates in astrophysical plasmas such as the core of stars or nucleosynthesis-relevant environments, extrapolations from accelerator-based experimental data to low energies are required \cite{Adelberger2011RMP, NegoitaRRP2016, CerjanJPG2018}. Direct measurements of reaction rates in laser-generated plasmas provide the chance for an alternative solution \cite{NegoitaRRP2016}. On the other hand, in plasmas, long-range electric fields are screened down by the dynamic flow of particles moving in response to electric fields. Owing to this charge screening effect, nuclear reactions could be drastically affected inside plasmas. The plasma screening effect for nuclear reactions has been intensively studied theoretically \cite{Salpeter1954AJP, Gruzinov1998APJ, Keller1953APJ, Salpeter1969APJ, Graboske1973APJ, Dzitko1995APJ, Bahcall2002AA, Chitanvis2007APJ, Shaviv1996APJ, Shaviv2000APJ, Mao2009APJ, KushnirMNRAS2019}, but remains an unsolved issue as experimental tests have not been performed so far.

We note that neutron production by irradiating deuterated polystyrene or D$_2$ targets at ultrahigh intensity has been achieved experimentally and investigated theoretically \cite{NorreysPPCF1998, DisdierPRL1999, PretzlerPRE1998, IzumiPRE2002, ZulickAPL2013, ToupinPOP2001, HabaraPRE2004a, HabaraPRE2004b}. Laser pulses with ultrahigh intensity ($\sim 10^{19}$ W/cm$^2$ or even higher) have been used, leading to the neutron production mechanism of the ion beam-target interaction, in which the ion beam is generated by the laser-target interaction then produces neutrons through the ion beam-target interaction. In contrast, here we focus on a different regime of relatively low laser intensity ($\leqslant 10^{18}$ W/cm$^2$) to avoid contributions of the beam-target interaction from energetic ions. Moreover, with Petawatt-class lasers and low intensity, we can create plasmas lasting for a timescale longer than the electron-ion equilibration time. Under this condition, the plasma can achieve thermal equilibrium, leading to thermonuclear reactions.

We first study the neutron events as functions of the plasma temperature assuming a general spherical plasma model, and calculate the neutron spectra in Sec.~\ref{sec:sph}. Then we model the plasma formation by the particle in cell (PIC) method for CD$_2$ solid-state targets in Sec.~\ref{sec:hdplas}. Plasma screening effect for thermonuclear reactions is discussed in Sec.~\ref{sec:screen} to show the astrophysical interests of the study. Then we conclude the paper with a brief summary in Sec.~\ref{sec:con}. We use the centimetre-gram-second system of units with $k_b = 1$, unless for some quantities the units are explicitly given.

%%%--------------------------------------theory and results-------------------------------------------------------
%%%--------------------------------------spherical plasmas-------------------------------------------------------
\section{Neutron production in spherical plasmas \label{sec:sph}}

Reactions of interests can be expressed as $n_1 + n_2 \rightarrow n_3 + n_4$, where $n_k$ stands for the number of the nuclear species $k$. The reaction rate per unit volume is \cite{Xu2013NPA} 
\begin{equation}
  P_{12} = \rho N_{A} \Lambda_{12 \rightarrow 34}, 
\end{equation}
where $\rho$ is the matter density, $N_A$ is Avogadro number, and  $\Lambda_{12 \rightarrow 34} = N_A <\!\! \sigma v \!\!>_{12} Y_1 Y_2 \rho/(n_1 !)$, with $Y_k = X_k/A_k$ ($X_k$ is the mass fraction and $A_k$ is the mass number), $\sigma$ the nuclear reaction cross section, and $v$ the relative velocity of the reactants. Then the total event number $N_t$ is 
\begin{equation}
  N_t = \int dV dt P_{12}.
\end{equation}
Assuming a uniform plasma with a spherical shape, the total event number is $N_t = P_{12} V \tau_p$ with the plasma volume $V$ and the plasma lifetime $\tau_p$. The plasma lifetime can be provided by the hydrodynamic expansion \cite{Krainov2002PR, Gunst2015POP},
\begin{equation} \label{eq:taup}
  \tau_p = R_p \sqrt{m_{\rm{ion}}/(T_e Z_{\rm{ion}})},
\end{equation}
where $R_p$ is the plasma radius, $m_{\rm{ion}}$ is the ion mass, $T_e$ is the plasma temperature [in Eq.~(\ref{eq:taup}), $T_e$ is in units of erg], and $Z_{\rm{ion}}$ is the ion charge state.

%%%%%%%%%%%%%%%%%%%%%%%%%%%%%%%%%%%%%%%%%%%
\begin{figure}[!h]
  \begin{center}
  %\flushleft
    \includegraphics[width=\columnwidth]{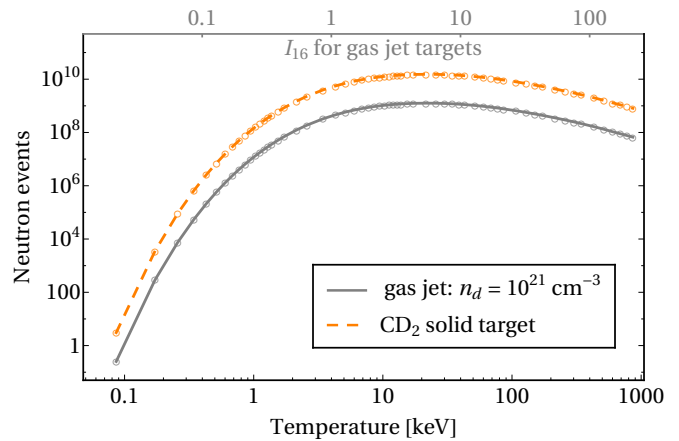}
  \end{center}
\caption{Neutron events as functions of the temperature in plasmas generated by D$_2$ gas jet targets and CD$_2$ solid-state targets. A laser pulse with energy $E_{\rm{laser}} = 500$ J is assumed, and carbon and deuterium ions are assumed to be fully ionized. For gas jet targets, a laser absorption fraction $f_{\rm{abs}} = 0.1$ is assumed. For CD$_2$ solid-state targets, a laser absorption fraction $f_{\rm{abs}} = 0.2$ is assumed and the carbon number density of the target is $4 \times 10^{22}$ cm$^{-3}$. The top frame shows the laser intensity to generate the plasma according to the scaling law Eq.~(\ref{eq:scal}), and it is valid for the gas jet case. $I_{16}$ is the laser intensity in units of $10^{16}$ W/cm$^2$. We note that, realistically, for the case of CD$_2$ solid-state targets the heating of the target is mainly conducted by hot electrons generated in the laser-target interaction, which overcomes the limit of direct heating by lasers above the critical density. PIC simulations are needed to model the plasma generation, which will be discussed in Sec.~\ref{sec:hdplas}. Here In order to understand the accessible range of neutron events as a function of the plasma temperature in the case of CD$_2$ solid-state targets, we assume at first the simple spherical plasma model.} \label{fig:neu_tn}
\end{figure}
%%%%%%%%%%%%%%%%%%%%%%%%%%%%%%%%%%%%%%%%%%% 

We first consider the reaction $^2$H($d$, $n$)$^3$He in plasmas generated by D$_2$ gas jet targets. The highest density of gas jets so far is around $10^{21}$ cm$^{-3}$ \cite{SchmidRSI2012, SyllaRSI2012}. In this case, the plasma volume can be estimated by $V = E_{\rm{laser}} f_{\rm{abs}}/[T (n_{\rm{ion}} + n_e)]$, where $E_{\rm{laser}}$ is the laser pulse energy and $f_{\rm{abs}}$ is the laser absorption fraction. The number of neutron events per laser shot as a function of the plasma temperature is shown in Fig.~\ref{fig:neu_tn}, for the case of the deuterium number density $n_d = 10^{21}$ cm$^{-3}$.  A laser pulse with energy $E_{l\rm{aser}} = 500$ J and the laser absorption fraction $f_{\rm{abs}} = 0.1$ are assumed. Reaction rates of $^2$H($d$, $n$)$^3$He are taken from the NACRE II database \cite{Xu2013NPA}. As shown in Fig.~\ref{fig:neu_tn}, the neutron events reach a maximal value of approximately $10^9$ per pulse at a temperature of approximately $10$ keV.

For low density cases, i.e., plasmas generated by D$_2$ gas jet targets, following Refs.~\cite{Wu2018PRL, Gunst2018PRE}, we connect the laser parameter to the electron temperature by the scaling law
\begin{equation} \label{eq:scal}
  T_e \sim 3.6 I_{16} \lambda_{\mu}^2 ~ \rm{keV},
\end{equation}
where $I_{16}$ is the laser intensity in units of $10^{16}$ W/cm$^2$ and $\lambda_{\mu}$ is the wavelength in microns \cite{Brunel1986PRL, Bonnaud1991LPB, Gibbon1996PPCF}. The electron-ion equilibration time \cite{DitmirePRA1996, LifschitzBook1981} is $\sim 100$ ps for the temperature of a few keV. With the high power laser and relatively low intensity, the plasma lifetime is more than $500$ ps. Therefore, the plasmas under consideration last long enough to reach thermal equilibrium, which is rare in laser driven platforms. We thus assume thermal equilibrium to model the outcome of the thermonuclear reaction. The result of neutron events from the reaction $^2$H($d$, $n$)$^3$He in plasmas generated by D$_2$ gas jet targets is shown in Fig.~\ref{fig:neu_tn} (see the top frame). A laser wavelength $1.053$ $\mu$m is assumed. It is shown that the laser intensity at approximately $10^{16}$ W/cm$^2$ is favourable for neutron productions. We note that the scaling described by Eq.~(\ref{eq:scal}) is limited to the non-relativistic regime, hence the prediction for high laser intensities ($\sim 10^{18}$ W/cm$^2$) is not precise. However, we focus on the temperature of a few keV (corresponding to the laser intensity in the order of $10^{16}$ W/cm$^2$); the result for high laser intensities is shown for the sake of comparison. We note also that in order to heat spherical plasmas directly by lasers, the density of the plasma cannot excess the laser's critical density, i.e., $\approx 10^{21}$ cm$^{-3}$  for lasers with $1.053$ $\mu$m wavelength.

We consider also CD$_2$ solid-state targets. The high-density case is of particular interests also in astrophysics, since many interesting aspects of astrophysics involve very high densities ($\sim 10^{26}$ cm$^{-3}$), e.g., the core of a star where nuclear reactions play important roles for the evolution of the star as well as the nucleosynthesis. We note that in the high density regime, plasmas cannot be heated directly by lasers as discussed above. In this regime the heating of the target is mainly conducted by hot electrons generated in the laser-target interaction which overcomes the limit of direct heating by lasers above the critical density, and experiments and simulations have shown that it is possible to heat targets at the solid-state density to temperatures of a few hundreds eV or even a few keV \cite{Saemann1999PRL, Audebert2002PRL, Sentoku2007POP, Wu2018PRL}. PIC simulations are needed to model the plasma generation, which will be discussed in Sec.~\ref{sec:hdplas}. Here, in order to understand the accessible range of the neutron events as a function of the plasma temperature in the case of CD$_2$ solid-state targets, we assume at first the simple spherical plasma model to obtain the neutron events, which are shown also in Fig.~\ref{fig:neu_tn}. The result shows that the neutron events reach a maximal value of approximately $10^{10}$ per pulse at a temperature of approximately $10$ keV. 

We note that the angular distribution of neutrons produced from thermonuclear reactions is isotropic. If one uses this neutron source as a neutron beam with a certain direction, then only part of the neutrons can be used. However, neutron beams produced via laser-induced particle acceleration leading to high-energy particle beams that subsequently interact with a secondary target have also a quite large angle divergence or even are isotropic \cite{HigginsonPRL2015, Roth2013PRL, Pomerantz2014PRL}. The state-of-the-art laser-driven neutron beams reach a maximum intensity of $10^{10}$ $n$/sr in the direction of the ion beam, with intensities in other directions less than half of the maximum value \cite{Roth2013PRL}. Thus, the intensity of the neutron beams reached here is about one order of magnitude lower than the one of the state-of-the-art laser-driven neutron beams. As also shown in Fig.~\ref{fig:neu_tn}, a significant number of events can be achieved at temperatures of few hundreds eV for both cases of D$_2$ gas jet targets and CD$_2$ solid-state targets, and a measurable number of events can be achieved even at temperatures of $\sim 100$ eV for the case of CD$_2$ solid-state targets. This may allow us to make direct measurements of reaction rates at low temperatures of astrophysical interests \cite{NegoitaRRP2016, Adelberger2011RMP, Xu2013NPA}.

%%%%%%%%%%%%%%%%%%%%%%%%%%%%%%%%%%%%%%%%%%%
\begin{figure}[!h]
  \begin{center}
    \includegraphics[width=\columnwidth]{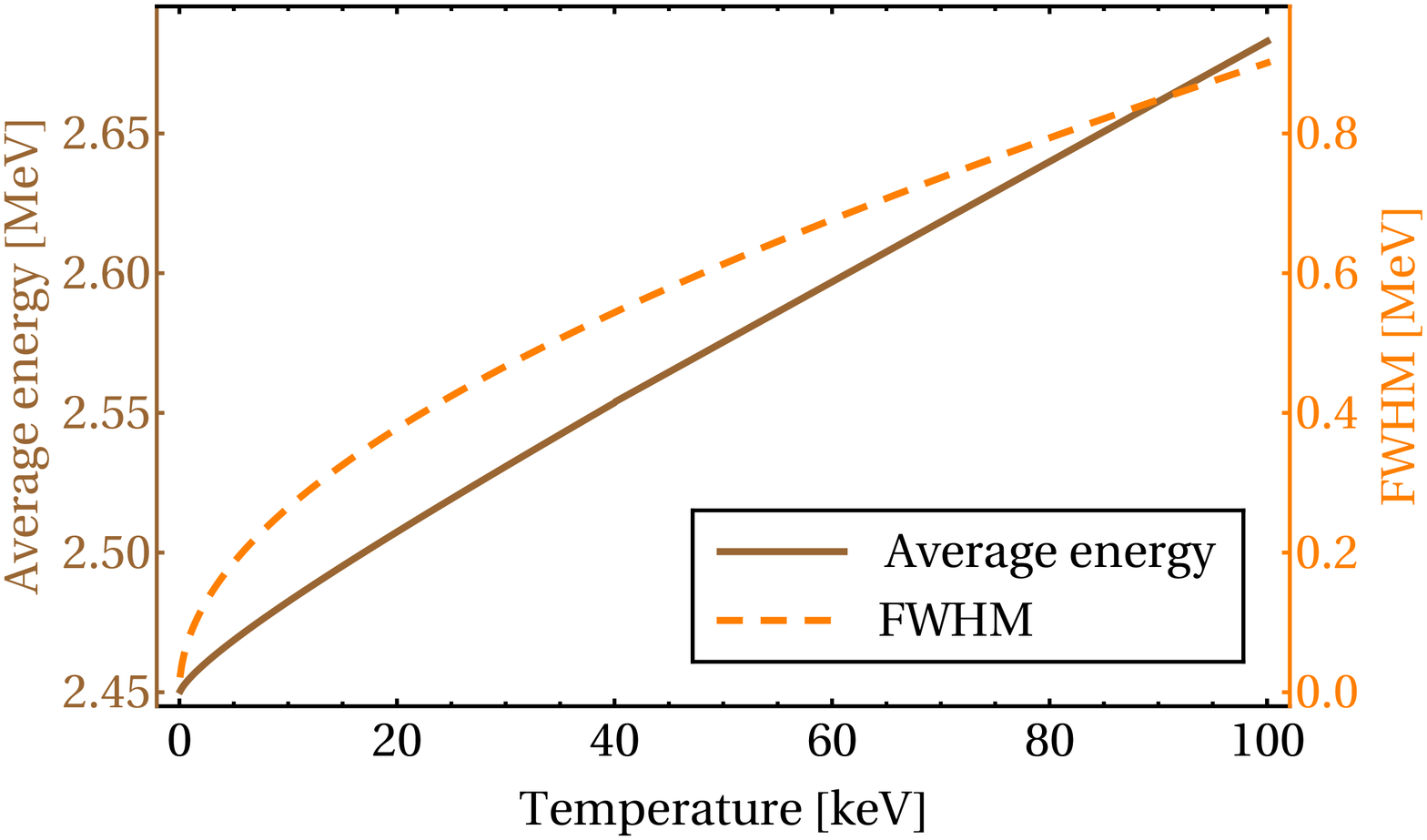}
  \end{center}
\caption{Average energy (brown solid curve; y-axis: left frame) and FWHM (orange dashed curve; y-axis: right frame) of neutrons produced from the reaction $^2$H($d$, $n$)$^3$He in plasmas as functions of the plasma temperature.} \label{fig:neu_enfwhm}
\end{figure}
%%%%%%%%%%%%%%%%%%%%%%%%%%%%%%%%%%%%%%%%%%%

In order to calculate the neutron spectra, we follow the relativistic calculation of fusion product spectra for thermonuclear reactions introduced in Refs.~\cite{Ballabio1997RSI, Ballabio1998NF}. The average neutron energy and FWHM of the neutron spectrum for the reaction $^2$H($d$, $n$)$^3$He as functions of the plasma temperature are shown in Fig.~\ref{fig:neu_enfwhm}. It is shown that for a plasma with a temperature of a few keV, the FWHM of the neutron spectrum is in the order of $100$ keV. This is about two orders of magnitude narrower than the one of the state-of-the-art laser-driven neutron beams \cite{Roth2013PRL}, which is in the order of $10$ MeV or even higher \cite{Roth2013PRL}.

%%%--------------------------------------high-density plasmas-------------------------------------------------------
\section{Neutron production in high-density plasmas generated by lasers \label{sec:hdplas}}

We now turn to the generation of plasmas in laser experiments in the high density case. Experiments and simulations have shown that it is possible to heat targets at the solid-state density to temperatures of a few hundreds eV or even a few keV \cite{Saemann1999PRL, Audebert2002PRL, Sentoku2007POP, Wu2018PRL}. Since in this regime the heating of the target is mainly conducted by secondary particles, i.e., hot electrons generated in the laser-target interaction, a more sophisticated model is necessary compared to the low-density case. We have performed a 2-dimensional (2-D) PIC simulation using the EPOCH code \cite{ArberPPCF2015}. A CD$_2$ target is assumed, with a carbon number density of $4 \times 10^{22}$ cm$^{-3}$, a thickness of $1$ $\mu$m ($x$-axis) and a length of $5$ $\mu$m ($y$-axis). The laser propagates along the $x$-axis, with a Gaussian profile in time with a FWHM of $100$ fs and a Gaussian profile in the $y$-axis centring at the center of the target with a FWHM of $2$ $\mu$m. The peak intensity and wavelength of the laser are $10^{18}$ W/cm$^2$ and $1.053$ $\mu$m, respectively. The size of the simulation box is $4$ $\mu$m $\times 6$ $\mu$m, and the target is located at the center of the simulation box. A rigid mesh with $800 \times 1200$ cells is used. The time step is $1.12 \times 10^{-17}$ s. A linear preplasma with thickness $0.5$ $\mu$m is considered in front of the target (the preplasma depends on the interaction of the prepulse of the laser with the target, and as a representative order, the preplasma assumed here is based on a similar ratio of the preplasma length to the target thickness in Ref.~\cite{Sentoku2007POP}). Carbon and deuterium ions are assumed to be fully ionized, and the numbers of pseudoparticles per cell are 50, 100, and 400 for carbon ions, deuterium ions, and electrons, respectively. 

\begin{figure}[!h]
  \begin{center}
    \includegraphics[width=0.49\columnwidth]{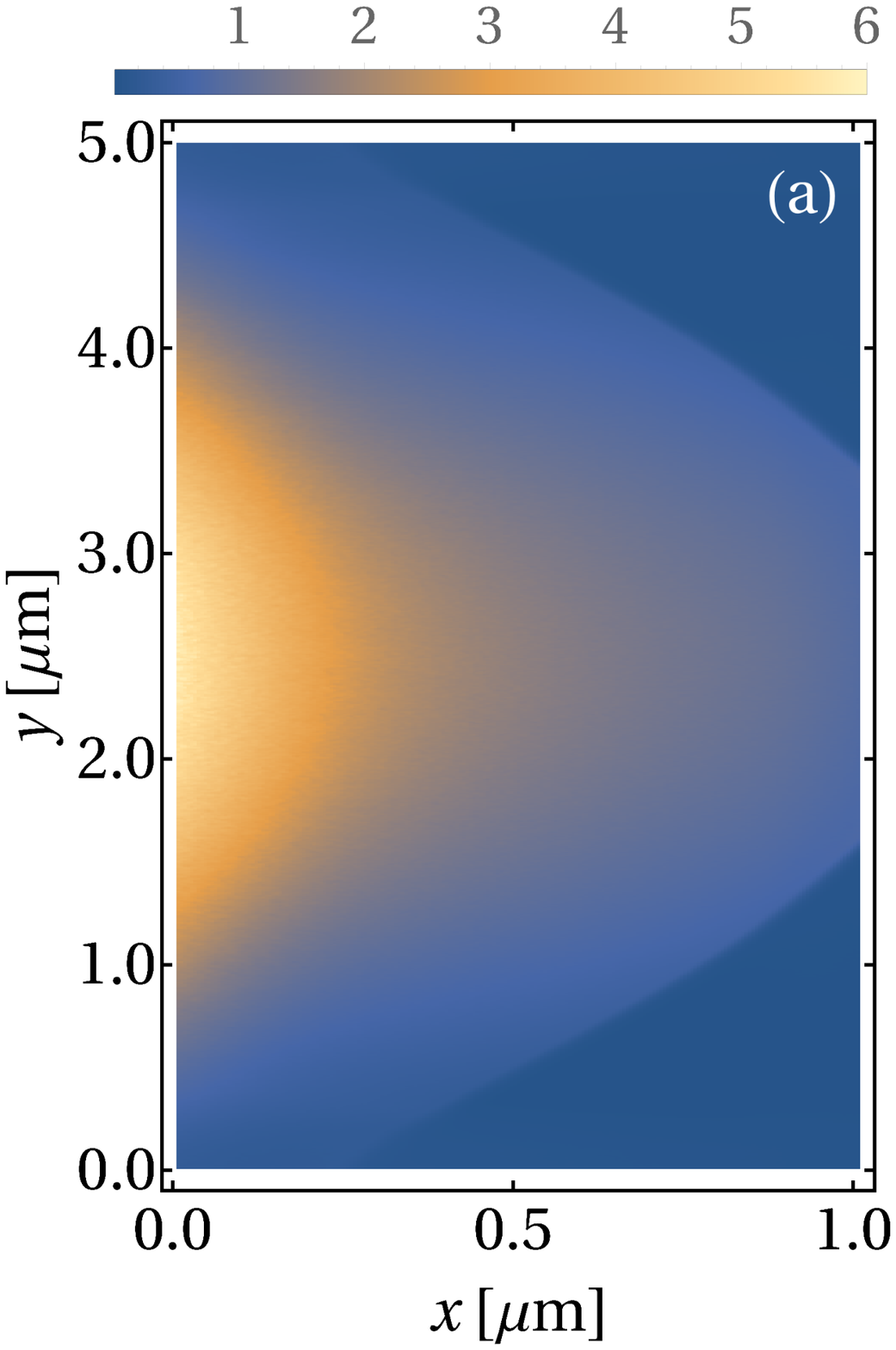}
    \includegraphics[width=0.49\columnwidth]{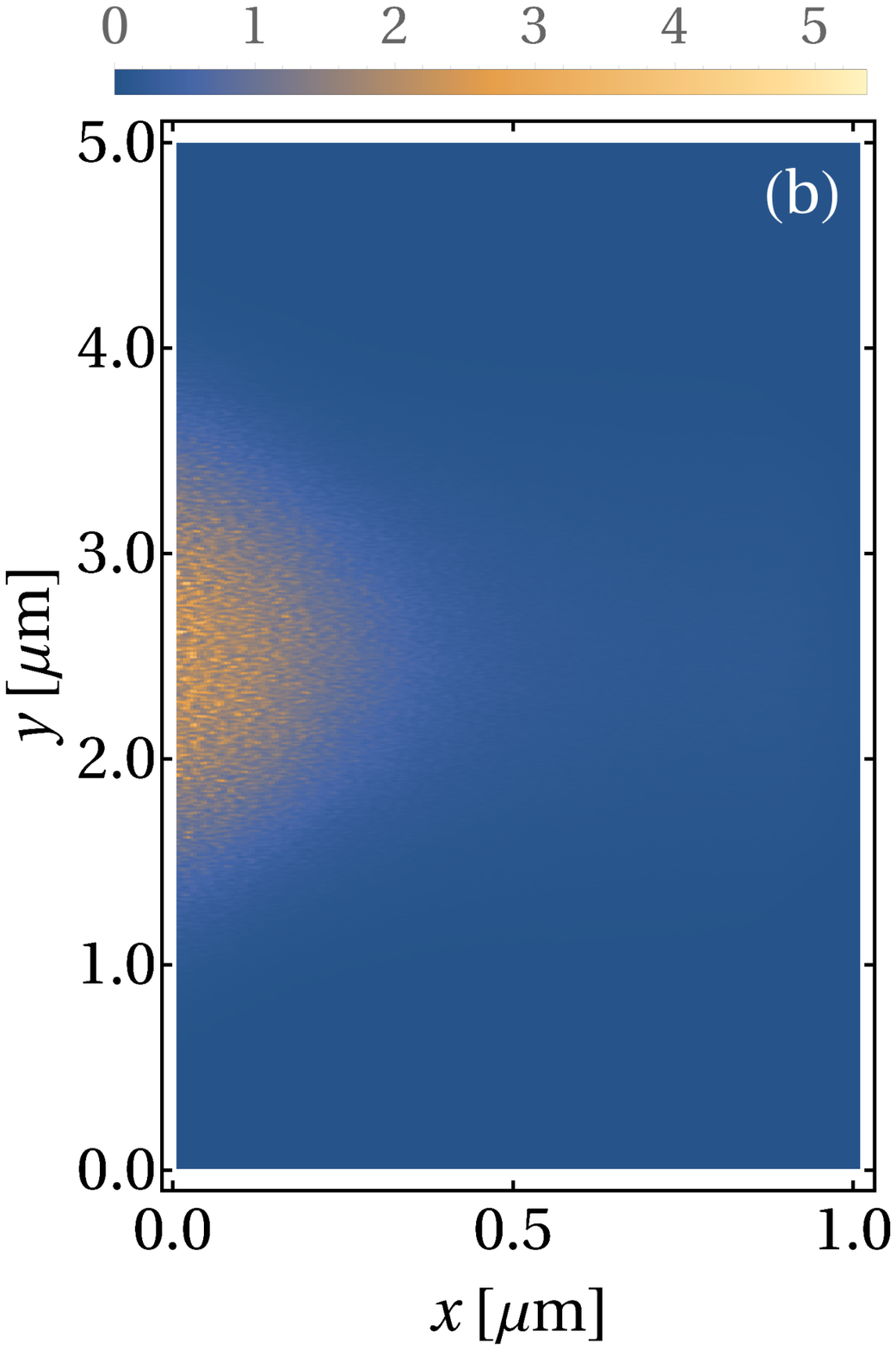}
    \includegraphics[width=0.99\columnwidth]{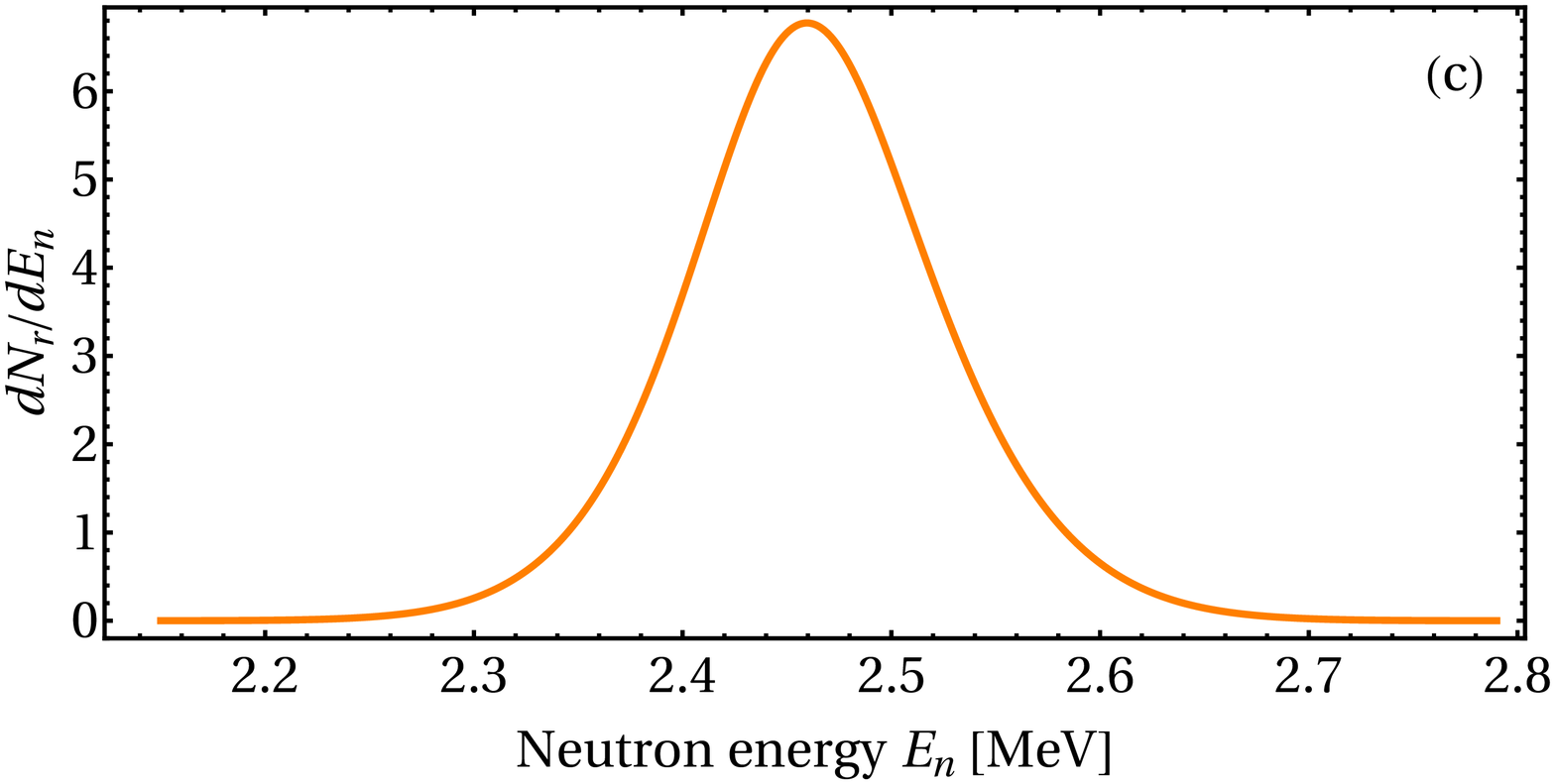}
  \end{center}
\caption{Results in the solid-target region of the 2-D PIC simulation at $2.5$ ps. (a): Temperature in units of keV. (b): Neutron production rate per unit volume of the reaction $^2$H($d$, $n$)$^3$He in units of $10^{25} $cm$^{-3}$ s$^{-1}$. (c): Normalised neutron distribution ($d N_r/d E_n$). A CD$_2$ target with a carbon number density of $4 \times 10^{22}$ cm$^{-3}$ is assumed, and carbon and deuterium ions are assumed to be fully ionized. The laser has a Gaussian profile in time with a FWHM of $100$ fs and a Gaussian profile in the $y$-axis with a FWHM of $2$ $\mu$m. The peak laser intensity is $10^{18}$ W/cm$^2$ and the laser wavelength is $1.053$ $\mu$m.} \label{fig:pic2d}
\end{figure}

The simulation and analysis have been performed until $2.5$ ps, and the electron temperature starts getting stable from $2.0$ ps. The electron temperature in the solid-target region at $2.5$ ps is shown in Fig.~\ref{fig:pic2d}(a). It is shown that the target can be heated to a few keV at the solid-state density in the laser focal spot. With such density and temperature, we obtain the neutron production rate per unit volume of the reaction $^2$H($d$, $n$)$^3$He, shown in Fig.~\ref{fig:pic2d}(b). We assume again thermal equilibrium, based on the following analysis. For the considered solid-state density, the electron-ion equilibration time \cite{DitmirePRA1996, LifschitzBook1981} is approximatively $10$ ps for the temperature of a few keV. We note that the density of the plasma decreases after $2.5$ ps as the plasma expands. Following the hydrodynamic model in Ref.~\cite{Gunst2015POP},  the plasma expansion leads to a neutron production timescale of approximatively of $20$ ps, and the effect of the plasma expansion on collision rates in the timescale of approximatively $20$ ps of interaction time leads to a deviation of approximatively $20\%$. Due to radiative processes, the plasma is also cooling. Estimates by the collisional-radiative code FLYCHK \cite {ChungHEDP2005} show that the timescale of the radiative cooling of the plasma at a temperature of around $1$ keV with the solid-state density is approximatively $30$ ps. Thus, the plasma lasts long enough to reach thermal equilibrium. Therefore, the use of the energy density at $2.5$ ps for neutron production as an approximation is justified.

%%%%%%%%%%%%%%%%%%%%%%%%%%%%%%%%%%%%%%%%%%%
%%%%%%%%%%%%%%%%%%%%%%%%%%%%%%%%%%%%%%%%%%% 

The normalised neutron distribution $d N_r/d E_n$ (normalised by the total neutron number per pulse) from the whole solid-target region is shown in Fig.~\ref{fig:pic2d}(c) as a function of the neutron energy $E_n$. It is shown that the energy bandwidth of the neutron beam is approximatively $100$ keV, which is about two orders of magnitude narrower than the one of the state-of-the-art laser-driven neutron beams \cite{Roth2013PRL}.

The 2-D PIC simulation predicts a conversion efficiency from laser energy to neutron yield of $10^6$ $n$/J. We note that, in order to exactly model the physical system, 3-D PIC simulations with the exact size of the target and laser focal spot are required, which are however impossible for Petawatt-class lasers with an intensity of $10^{18}$ W/cm$^2$ because of a lack of computing power. In order to model the plasma, a 2-D PIC simulation has been performed here. Obviously, the 2-D simulation cannot exactly describe the physical system since it misses information on one of the dimensions. In order to check the possible influence of laser intensity variation in the laser focal spot, we have also performed a 1-D PIC simulation, which in principle could be applied for any size of the laser focal spot by completely neglecting the information on the latter. The 1-D PIC simulation predicts similar results as the 2-D PIC simulation in the region of the laser focal spot. This indicates that edge effects, losses, and gradients, which all very much depend on the actual geometry in a real scenario, will not lead to a dramatic change for the neutron production case. Therefore the insight gained on a simplified scale in our 2-D PIC simulation may imply a similar conversion efficiency from laser energy to neutron yield for a certain amount of driver energy on the scale of $500$ J for the current state-of-the-art laser systems. This would lead to the generation of a neutron beam with approximatively $10^9$ neutrons per pulse with a laser pulse energy of $500$ J. The intensity of such neutron beam is about $2$ orders of magnitude lower than the one of the state-of-the-art laser-driven neutron beams (maximum intensity $10^{10}$ $n$/sr) \cite{Roth2013PRL}.

As discussed in many experiments of neutron productions by irradiating deuterated polystyrene or D$_2$ targets at ultrahigh intensity, the ion beam-target interaction is responsible for neutron productions. As shown in Ref.~\cite{ToupinPOP2001}, the contribution from the ion beam-target interaction is suppressed for the relatively low laser intensity under consideration ($\leqslant 10^{18}$ W/cm$^2$). Re-scaling the laser energy to our case, neutron events from the beam-target interaction [$^2$H($d$, $n$)$^3$He] are more than $3$ orders of magnitude less than our results. Other possible neutron sources from the reactions $^2$H($d$, $pn$)$^2$H, $^2$H($e$, $en$)$p$, $^2$H($\gamma$, $n$)$p$, and $^{12}$C($d$, $n$)$^{13}$N are negligible \cite{ToupinPOP2001} in our case. We note that neutron beams produced by thermonuclear reactions start delayed due to the required electron-ion equilibration time when compared to neutrons driven by laser ion acceleration concepts.

%%%--------------------------------------screening effects-------------------------------------------------------
\section{Plasma screening effects \label{sec:screen}}

\begin{figure}[!h]
  \begin{center}
    \includegraphics[width=\columnwidth]{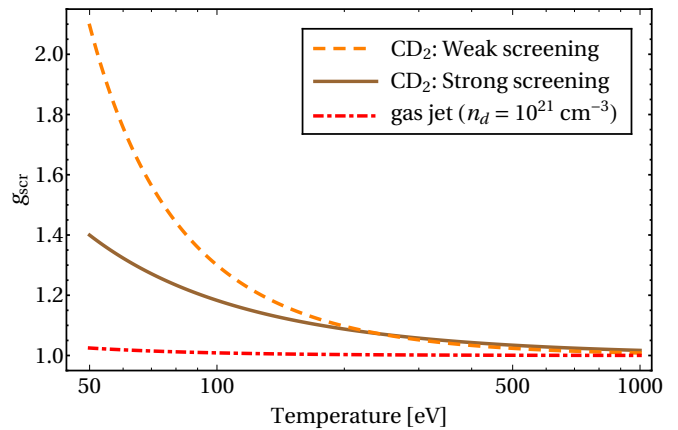} 
  \end{center}
\caption{Plasma screening enhancement factor for the reaction $^2$H($d$, $n$)$^3$He as functions of the plasma temperature, in plasmas generated by D$_2$ gas jets with $n_d = 10^{21}$ cm$^{-3}$ (weak-screening: red dot-dashed curve), and in plasmas generated by CD$_2$ solid-state targets (weak-screening: orange dashed curve; strong-screening: brown solid curve). Carbon and deuterium ions are assumed to be fully ionized. The carbon number density of CD$_2$ solid-state targets is $4 \times 10^{22}$ cm$^{-3}$.} \label{fig:gscreen}
\end{figure}

As a further advantage, we analyze the plasma screening effect for thermonuclear reactions \cite{Adelberger2011RMP, NegoitaRRP2016, CerjanJPG2018, Wu2017APJ}. Due to the plasma screening effect, the reaction rate can be enhanced by a factor \cite{AtzeniBook2004}, $<\!\!\! \sigma v\!\!\!>_{\rm{scr}} = g_{\rm{scr}} \!\! <\!\!\! \sigma v \!\!\!>$. In weakly coupled plasmas, the screening enhancement factor is \cite{Salpeter1954AJP, Gruzinov1998APJ, Wu2017APJ} 
\begin{equation}
  g_{\rm{scr}} = \exp{\left[ Z_1 Z_2 \alpha/(T \lambda_{\rm{D}}) \right]}, 
\end{equation}
with the Debye length $\lambda_{\rm{D}}$, the fine-structure constant $\alpha$, and nuclear charges of the reactants $Z_1$ and $Z_2$. The plasma screening enhancement factor for the reaction $^2$H($d$, $n$)$^3$He is shown in Fig.~\ref{fig:gscreen}. It is shown that the plasma screening effect for the reaction $^2$H($d$, $n$)$^3$He in plasmas generated by D$_2$ gas jets is negligible. However, the plasma screening effect is greatly enhanced in the case of CD$_2$ solid-state targets. As neutron events are measurable for low temperatures shown in Fig.~\ref{fig:neu_tn}, it may lead to the possibility of direct measurements of the plasma screening effect for thermonuclear reactions.

%%%%%%%%%%%%%%%%%%%%%%%%%%%%%%%%%%%%%%%%%%%
%%%%%%%%%%%%%%%%%%%%%%%%%%%%%%%%%%%%%%%%%%% 

Furthermore, low temperature plasmas generated by CD$_2$ targets may reach the condition for strong screening \cite{Salpeter1954AJP}, $Z_1 < \rho^{1/3}$ and $0.23 Z_1^{2/3} z (\xi\rho)^{1/3} T_6^{-1} > 1$ (for $T \sim 100$ eV, $0.23 Z_1^{2/3} z (\xi\rho)^{1/3} T_6^{-1} \sim 1$). Here $T_6$ is the plasma temperature in units of $10^6$ K and $\xi = \sum_{i} (X_i Z_i)/A_i$. In this case, the plasma screening enhancement factor is 
\begin{equation}
  g_{\rm{scr}} = \exp{\! \left\{ 0.205 \! \left[(Z_1 \!+\! Z_2)^{5/3} \!-\! Z_1^{5/3} \!-\! Z_2^{5/3}\right] \! (\xi \rho)^{1/3} T_6^{-1} \right\}}. 
\end{equation}
The strong screening enhancement factor for the case of CD$_2$ solid targets is shown in Fig.~\ref{fig:gscreen}. For temperatures of $\sim 100$ eV, the strong screening enhancement factor is significantly different from the weak screening one, which could lead to the experimental determination of the strong screening effect for fusion reactions in plasmas.

%%%--------------------------------------summary-------------------------------------------------------
\section{Summary \label{sec:con}}

In conclusion, we have studied the production of intense neutron beams via thermonuclear reactions in laser-generated plasmas. The reaction $^2$H($d$, $n$)$^3$He in plasmas generated by Petawatt-class lasers interacting with D$_2$ gas jet targets and CD$_2$ solid-state targets has been analyzed. Intense neutron beams of about two orders of magnitude narrower bandwidth can be obtained from thermonuclear reactions, as compared to the state-of-the-art laser-driven neutron beams \cite{Roth2013PRL, Pomerantz2014PRL}. The intensity of such neutron beams is about one or two orders of magnitude lower than the one of the state-of-the-art laser-driven neutron beams, which indicates that the spectral brightness is similarly high or even higher than the one of the state-of-the-art laser-driven neutron beams. Such neutron beams with narrow bandwidth have numerous advantages in the applications in industry and fundamental research, such as in the interrogation of material and life science \cite{ZaccaiScience2000, MaNatNano2013} using a neutron beam, and neutron capture experiments for fundamental nuclear physics and nuclear astrophysics \cite{BleuelPFR2016}. Compared to the state-of-the-art laser-driven neutron beams with a broad bandwidth, narrow bandwidth of the beam could lead to more clear and better understandable signal, as it avoids a large divergence and uncertainty in the probe energy (for the interrogation) or the reactant energy (for the neutron capture). Furthermore, for applications involved resonance processes, the spectral brightness is important for the number of the events that can be obtained.

We have also pointed out the possible astrophysical implications of our work, i.e., direct measurements of reaction rates at low temperatures of nucleosynthesis-relevant energies, and the great enhancement on the plasma screening effect which may open new possibilities to study this so far open issue in astrophysics.

%%%%%%%%%%%%%%%%%%%%%%%%%%%%%%%%%%%%%%%%%%%%%%%%%%%%%%%%%

\begin{acknowledgements}
  We thank Adriana P\'alffy and Christoph H. Keitel for fruitful discussions.
\end{acknowledgements}

\bibliography{neuprefs}{}

\end{document}